\documentclass[prb,twocolumn,showpacs,preprintnumbers,amsmath,amssymb]{revtex4}

\usepackage{graphicx}
\usepackage{dcolumn}
\usepackage{bm}
\usepackage{epsfig}
\usepackage{graphicx}

\usepackage{hyperref}
\hypersetup{
    colorlinks,%
    citecolor=blue,%
    filecolor=black,%
    linkcolor=blue,%
    urlcolor=black
}

\def\HG#1 {\emph{\color{red}#1} }

\begin{document}

\title{Crystal growth and intrinsic magnetic behavior of Sr$_2$IrO$_4$}

\author{N. H. Sung, H. Gretarsson, D. Proepper, J. Porras, M. Le Tacon, A. V. Boris, B. Keimer and B. J. Kim}
\affiliation{Max-Planck-Institut f\"{u}r Festk\"{o}rperforschung, Heisenbergstr. 1, D-70569 Stuttgart, Germany
}

\begin{abstract}
We report on the growth of stoichiometric Sr$_2$IrO$_4$ single crystals, which allow us to unveil their intrinsic magnetic properties.
The effect of different growth conditions has been investigated for crystals grown by the flux method.
We find that the magnetic response depends very sensitively on the details of the growth conditions.
We assess the defect concentration based on magnetization, X-ray diffraction, Raman scattering, and optical conductivity measurements.
We find that samples with a low concentration of electronically active defects show much reduced in-gap spectral weight in the optical conductivity and a pronounced two-magnon peak in the Raman scattering spectrum.
A prolonged exposure at high temperature during the growth leads to higher defect concentration likely due to creation of oxygen vacancies.
We further demonstrate a systematic intergrowth of Sr$_2$IrO$_4$ and Sr$_3$Ir$_2$O$_7$ phases by varying the growth temperature.
Our results thus emphasize that revealing the intrinsic magnetic properties of Sr$_2$IrO$_4$ and related materials requires a scrupulous control of the crystal growth process.

\end{abstract}
\maketitle

\section{Introduction}

\normalsize

The interest in iridates grows fueled by the prospect of discovering novel electronic and/or topological orders that emerge from the interplay of spin-orbit coupling and electron correlations.
Sr$_2$IrO$_4$ is an archetypal example of such new phases, exhibiting a Mott gap and a distinct spin-orbital entangled magnetic structure \cite{BJKim2008, BJKim_Science_2009}.
Although Sr$_2$IrO$_4$ has been studied since the early 90's in connection to high temperature superconducting cuprates \cite{PhysRevB.49.9198, Cava1994}, the important role of spin-orbit coupling had not been appreciated then.
From recent studies of electronic and magnetic excitation spectra, it is now well understood that the insulating nature of Sr$_2$IrO$_4$ is due to strong spin-orbit coupling \cite{BJKim2008}, suppression of which results in a metallic phase in Sr$_2$RhO$_4$ \cite{Cao2012Rh}, a 4$d$ counterpart of Sr$_2$IrO$_4$.

The magnetic structure and dynamics of Sr$_2$IrO$_4$ are now reasonably well understood within the theoretical framework in the strong spin-orbit coupling limit described in terms of the $j$-1/2 Kramers doublet.
Remarkably, the $j$-1/2 isospins allow rich expressions of magnetic coupling depending on the lattice and bonding geometry, ranging from isotropic Heisenberg interaction to  a bond-directional, Ising-like interaction \cite{GJGK2009}.
The latter provides a basis for implementing exotic quantum magnets inspired by the Kitaev model \cite{Kitaev}.
The former surprisingly indicates that a system with very strong spin-orbit coupling can mimic a system without it.
Experimentally, it has been confirmed that the canted antiferromagnetic structure of Sr$_2$IrO$_4$ derives from a predominantly isotropic nearest-neighbor magnetic exchange coupling \cite{JunghoKim2012, SFujiyama, JGVale}.
Thus, Sr$_2$IrO$_4$ becomes the first non-cuprate material to realize a (iso)spin-1/2 Heisenberg antiferromagnet in a quasi-two-dimensional square lattice---a possible platform for high temperature superconductivity \cite{ZYMeng, PhysRevLett.106.136402, PhysRevB.89.094518, JKIM2014, Kim11072014}.

Most recently, tantalizing indications of $d$-wave superconductivity have been seen in electron-doped Sr$_2$IrO$_4$ \cite{arXiv:1506.06639,DLFeng}: a gap of predominantly $d$-wave symmetry is observed in the low-temperature angle-resolved photoemission spectra \cite{arXiv:1506.06639}, and a particle-hole symmetric density of states is observed in scanning tunneling spectra below a temperature as high as 50 K \cite{DLFeng}.
Above this temperature, the $d$-wave gap evolves into a pseudogap, in striking parallel with the cuprates \cite{Kim11072014, arXiv:1506.06639}.
However, because of the inherently surface sensitive nature of the technique used to dope the sample in these studies, involving in situ deposition of alkali metal atoms on the surface of undoped Sr$_2$IrO$_4$, it remains unclear at this point whether the $d$-wave gap originates from superconductivity without evidence from bulk measurements.
At least, these results indicate that highly nontrivial electronic phases emerge from doping a ``spin-orbit'' Mott insulator.

The more traditional method of chemical doping \cite{PhysRevB.84.100402, PRB82_115117, JAP109_07D906} through , for example, substitution of Sr by La, has proven much more difficult than in cuprates.
La-doped Sr$_2$IrO$_4$ single crystals have only become available in the last couple of years \cite{Fisher, Baumberger, PRB92_075125(2015)}, but angle-resolved photoemission studies on them \cite{Baumberger, Fisher} do not fully reproduce the observations made via the in situ doping method \cite{Kim11072014, arXiv:1506.06639}.
While it is unclear at the moment what leads to such differences between the two doping methods, we note that a well-defined quasiparticle peak is absent for any doping level in La-doped samples, indicating suppressed coherent motion of doped electrons presumably due to disorder created by chemical doping.

Moreover, we note that previous studies on the parent Sr$_2$IrO$_4$ and its sister compounds have shown significant variations in the magnetic properties even without doping, likely reflecting different defect concentrations \cite{PhysRevB.49.9198,PhysRevB.57.R11039,PhysRevB.80.140407,PhysRevB.84.100402,BJKim_Science_2009,Takagi,Fisher, JAP107_09D910, PRB92_075125(2015), PhysRevB.85.184432, PhysRevB.66.214412}.
For example, the sizable $c$-axis component of the uniform magnetic susceptibility is not understood and seems incompatible with the known magnetic structure of Sr$_2$IrO$_4$.
In addition, the diamagnetism initially reported for Sr$_3$Ir$_2$O$_7$ is completely unexplained and is absent in samples grown under different conditions \cite{PhysRevB.66.214412, PhysRevB.85.184432, Takagi}.
Therefore, we believe it is imperative to establish the intrinsic magnetic properties through growth of stoichiometric single crystals of the parent iridates before the doping-induced phenomena can be fully explored.

In this paper, we describe the synthesis and characterization of Sr$_2$IrO$_4$ single crystals.
Synthesis of stoichiometric single crystals was achieved using the flux method, and characterizations were performed using magnetization, x-ray diffraction, Raman scattering, and optical conductivity measurements.
We find that the magnetic response depends very sensitively on subtle details of the growth condition.
Through energy dispersive x-ray spectroscopy (EDX), optical spectroscopy, and Raman scattering, we infer that oxygen vacancies are the likely culprit of the divergent properties reported for Sr$_2$IrO$_4$.
Furthermore, by systematic variation of the growth temperature, we are able to introduce, in a controlled way, layers of Sr$_3$Ir$_2$O$_7$ into the bulk Sr$_2$IrO$_4$ single crystals. Our results thus emphasize that revealing the intrinsic magnetic properties of Sr$_2$IrO$_4$ and related materials requires careful attention the oxygen stoichiometry.

\section{Experimental Details}

Single crystals were grown using the flux method. Powders of IrO$_{2}$, SrCO$_{3}$, and SrCl$_{2}$$\cdot$6H$_2$O were thoroughly mixed and placed in a platinum crucible covered with a lid.
The crucibles were heated in a programmable box furnace in air then cooled to room temperature and removed from the furnace.
The details of the growth conditions are documented in table \ref{table1}-\ref{table3}.
After cooling, crystals were separated from the residual flux by rinsing out with distilled water. The crystals were found to have a regular shape with a large flat shiny surface with lengths up to few mm. The characterizations of crystals were performed using powder x-ray diffraction (XRD) in Debye-Scherrer geometry with Ag-K$\alpha$1 radiation, single crystal XRD in a high-resolution four-circle diffractometer with Cu-K$\alpha$1 radiation, as well as energy dispersive x-ray spectroscopy (EDX) equipped with a scanning electron microscope. The magnetization measurements were carried out using a commercial superconducting quantum interference device magnetometer (MPMS SQUID VSM, Quantum Design). The low temperature Raman scattering experiments were performed in the backscattering geometry using a JobinYvon LabRam HR800 spectrometer and the 632.8 nm excitation line of a HeNe laser was used.
The power of the laser beam was 1.5 mW and the diameter of the beam $\sim 10 \mu m$.
The samples were cooled down in a He-flow cryostat. To describe the scattering geometries, we use the Porto notation A(BC)A$^\prime$, where A(B) and A$^\prime$(C) stand for the propagation (polarization) directions of the incident and scattered light relative to the crystalline axes.
All of the data presented in this paper were taken with the light propagation direction along the crystalline $c$-axis; hence only the polarization orientations will be specified. Crystals were cleaved $ex$-situ to obtain a fresh clean surface.
The far infrared (FIR) ellipsometric measurements were carried out on a home-built ellipsometer attached to a Bruker 66v/S FT-IR spectrometer located at the IR1 beamline of the ANKA synchrotron at KIT, Germany. The experiments were performed on the same samples as used for the Raman spectroscopy studies. The complex optical conductivity $\sigma(\omega) = \sigma_1(\omega)+\text{i} \sigma_2(\omega)$ is obtained from the ellipsometric angles $\Psi(\omega)$ and $\Delta(\omega)$ by direct inversion of the Fresnel formulae.

\section{Result and Discussion}

\begin{figure}
\includegraphics[width=0.5\textwidth]{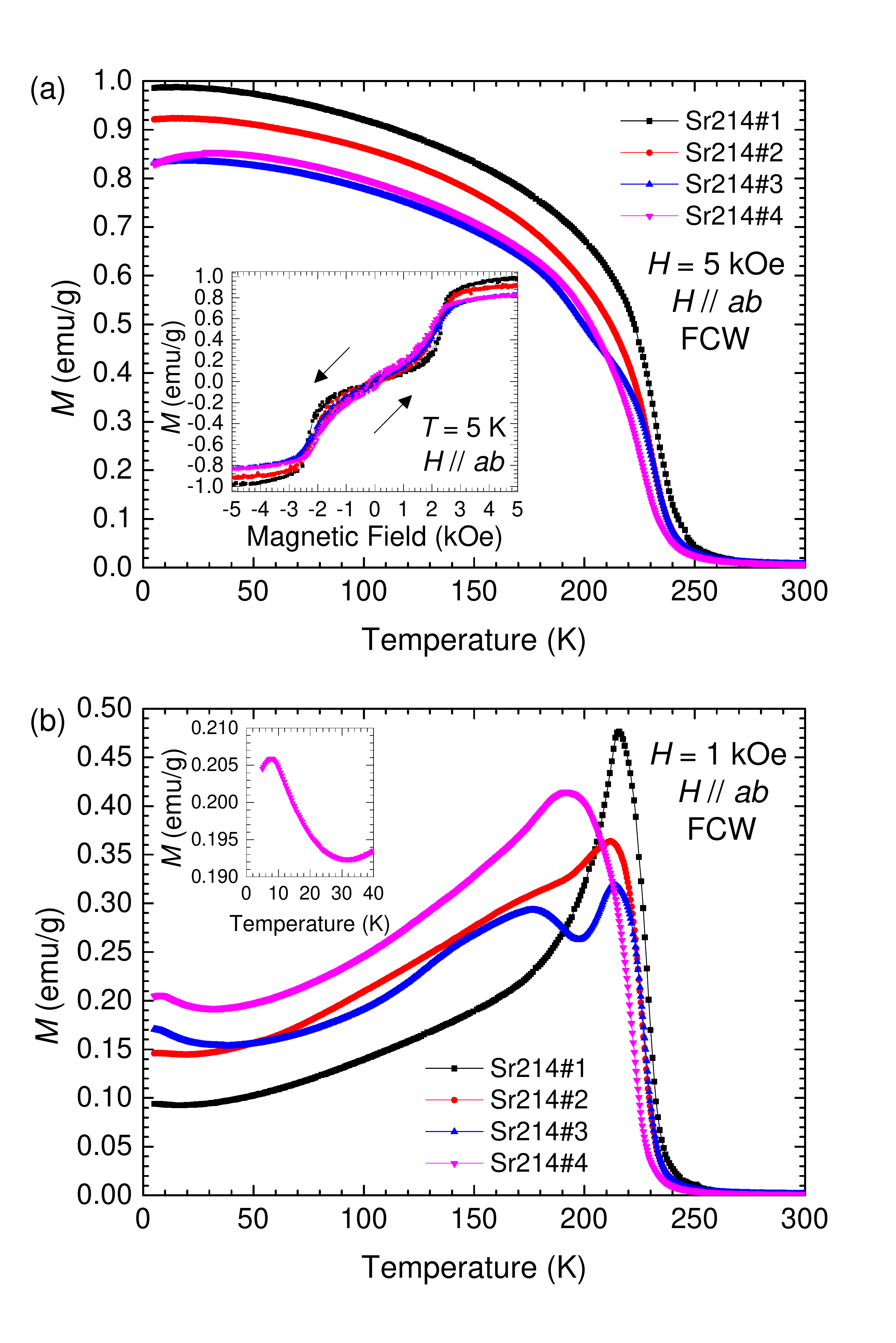}
\caption{
\label{FIGURE_F}(Color online)
 Temperature dependent magnetization, $M$($T$), of Sr$_2$IrO$_4$ single crystals measured on warming after field-cooling (FCW mode) with an applied field of (a) $H$ = 5 kOe and (b) $H$ = 1 kOe perpendicular to the crystallographic $c$-axis ($H$ $\parallel$ $ab$).
 The inset of (a) displays isothermal magnetization $M$($H$) of Sr$_2$IrO$_4$ single crystals at $T$ = 5 K ($H$ $\parallel$ $ab$) and the inset of (b) shows expended plot of $M$($T$) at low temperatures for Sr214$\#4$.}
\end{figure}

We start by looking at the magnetic properties of four Sr$_2$IrO$_4$ single crystals (labelled Sr214$\#1$ through $\#4$) grown from the same molar ratio of starting materials under varying heating sequences as described in table \ref{table1}.
The biggest difference in the growth condition is between Sr214$\#1$ and Sr214$\#4$.
While both start from heating the starting materials up to 1300 $^{\circ}$C, Sr214$\#4$ is exposed to elevated temperatures for a longer time.
The growth sequence of Sr214$\#2$ and Sr214$\#3$ are modifications of Sr214$\#1$ and Sr214$\#4$, respectively.
Note that our growth is performed in much lower temperature ranges than used in most of the earlier works \cite{Cao1480, PhysRevB.85.184432, PRB92_075125(2015)}.
In Fig.~\ref{FIGURE_F}(a), we plot the temperature dependence of the magnetization $M$($T$) of each crystal measured in a magnetic field of 5 kOe applied perpendicular to the crystallographic $c$-axis ($H$ $\parallel$ $ab$), well above the metamagnetic transition that occurs at $H_C$ $\approx$ 2 kOe (see inset of Fig.~\ref{FIGURE_F}(a)). Above this field, canted moments that result from rotation of oxygen octahedra stack ferromagnetically along the $c$-direction, and the uniform susceptibility looks similar to that of a ferromagnet.
Comparison of different curves shows that the highest saturated magnetic moment is achieved for Sr214$\#1$  ($M$ = 0.98 emu/g and 0.08 $\mu$$_B$/Ir at $T$ = 5 K, $H$ = 5 kOe). Other batches show significantly lower values ranging between 0.82 and 0.92 emu/g. Further, a `two-step' structure in the $M$($T$) curve is noticed for Sr214$\#3$, and a slight decrease in the magnetization is observed at low temperature below $\sim$30 K for Sr214$\#4$. In the inset of Fig.~\ref{FIGURE_F}(a), we plot the isothermal magnetization $M$($H$) curve of all samples. In Sr214$\#1$ the metamagnetic transition is much sharper compared to the other three crystals.

\begin{figure}
\includegraphics[width=0.5\textwidth]{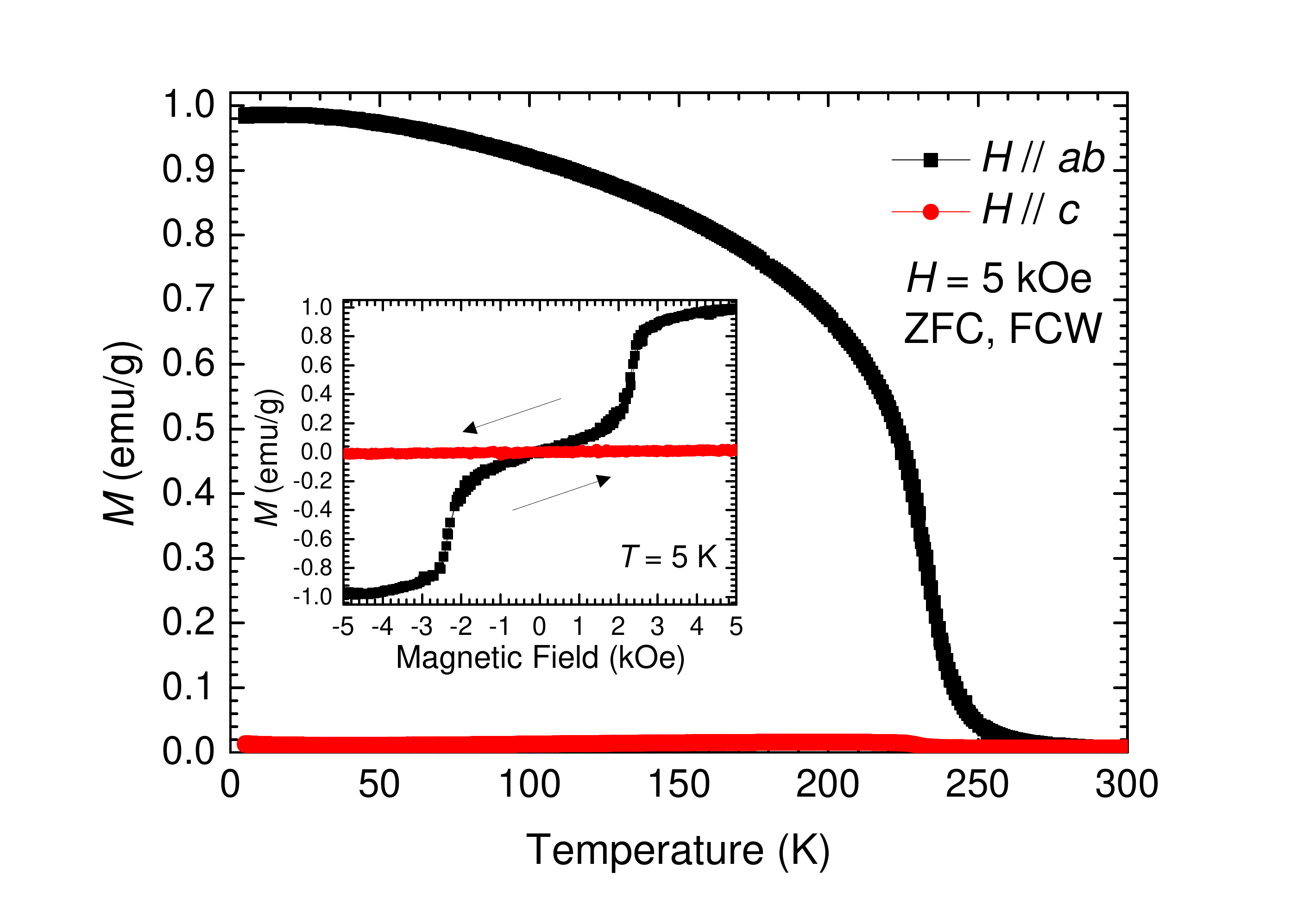}
\caption{
\label{FIGURE_B}
(Color online) Temperature dependence of magnetization $M$($T$) of a Sr$_2$IrO$_4$ single crystal (Sr214$\#1$) in zero field cooled (ZFC) and field cooled warming (FCW) modes at $H$ = 5 kOe with $H$ $\parallel$ $ab$ and $H$ $\parallel$ $c$, respectively. Inset: Isothermal magnetization $M$($H$) at $T$ = 5 K with $H$ $\parallel$ $ab$ and $H$ $\parallel$ $c$, respectively.
}
\end{figure}

These differences in the $M$($T$) curves (Fig.~\ref{FIGURE_F}(a)) are more prominent in measurements performed in applied field below the metamagnetic critical field at $H$ = 1 kOe, shown in Fig.~\ref{FIGURE_F}(b). Whereas Sr214$\#1$ with the highest saturated moment shows a sharp peak at $T$ = 215 K, Sr214$\#4$ shows a broad hump peaked at a much lower temperature $T$ = 175 K.
Two anomalies are found in Sr214$\#2$ and Sr214$\#3$, suggesting coexistence of the two features seen in Sr214$\#1$ and Sr214$\#4$.
It is noteworthy that previously published results share more resemblance to the  broad profile of Sr214$\#4$ rather than the sharp profile of Sr214$\#1$ \cite{PhysRevB.84.100402,PhysRevB.80.140407}. Sr214$\#4$ shows another anomaly in the low temperature region (inset of Fig.~\ref{FIGURE_F}(b)) similar to earlier data, based on which a transition to a different magnetic structure, possibly a non-coplanar structure that has a $c$-axis component, has been suggested \cite{PhysRevB.84.100402,PhysRevB.80.140407}. However, these anomalies are clearly absent in Sr214$\#1$.
Furthermore, we observe almost zero $c$-axis magnetic susceptibility (Fig.~\ref{FIGURE_B}), in sharp contrast to earlier studies that report a sizable $c$-axis response \cite{PhysRevB.57.R11039, PhysRevB.80.140407, PhysRevB.84.100402}.

Our measurements highlight the strong dependence of the magnetic properties of Sr$_2$IrO$_4$ to subtle differences in the growth condition.
As we shall see later on, we judge based on optical conductivity and Raman scattering spectra that Sr214$\#1$ has the lowest defect concentration among all crystals we have grown.
Therefore, we believe that the magnetic behavior reported above for Sr214$\#1$  is representative of the clean, ideal phase of Sr$_2$IrO$_4$.

\begin{figure}
\includegraphics[width=0.4\textwidth]{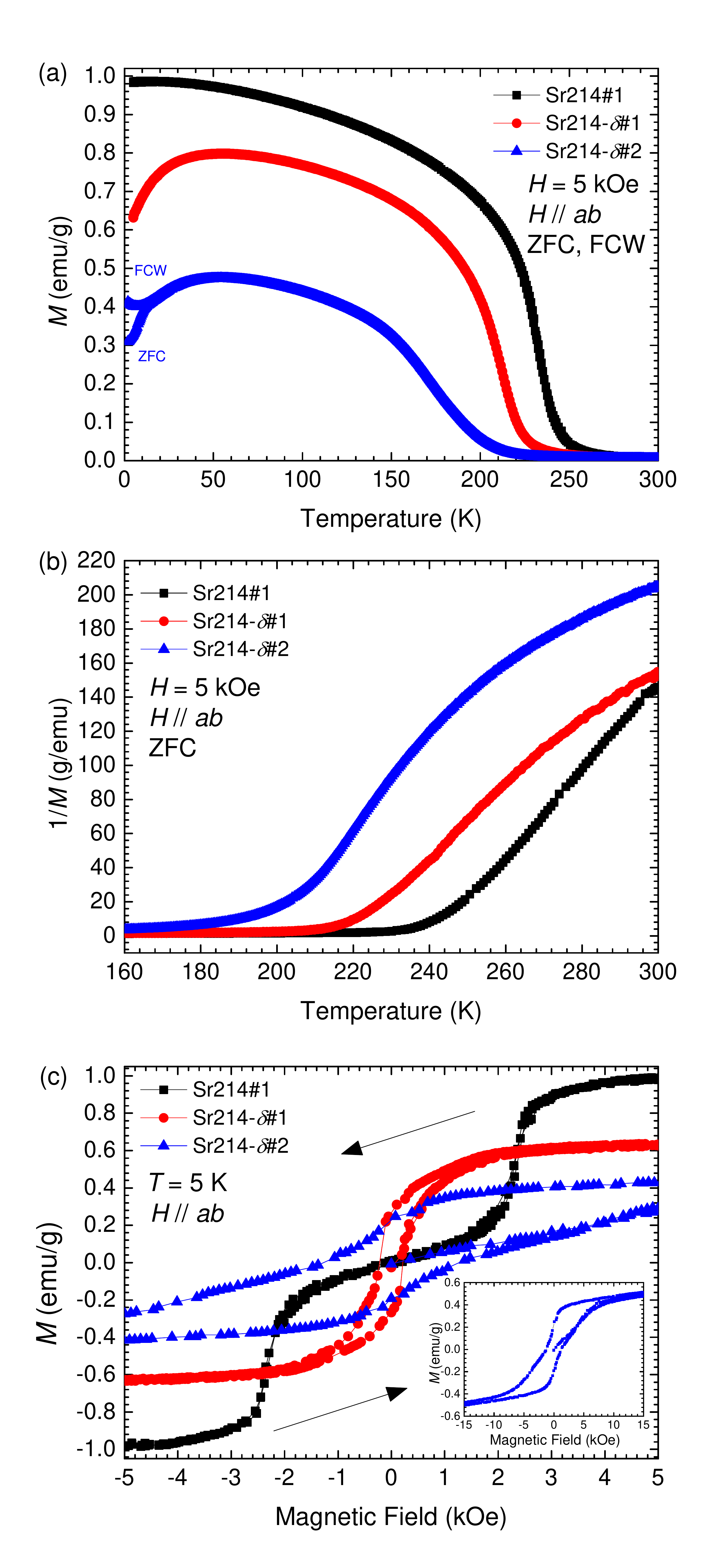}
\caption{
\label{FIGURE_AC}
(Color online) (a) Temperature dependence of magnetization $M$($T$) for Sr214$\#1$, Sr214$-\delta\#1$, and Sr214$-\delta\#2$ at $H$ = 5 kOe with $H$ $\parallel$ $ab$ in zero field cooled (ZFC) and field cooled warming (FCW) modes. (b) Inverse magnetic moments around $T_N$.
(c) Magnetic field-dependent magnetizations $M$($H$) for Sr214$\#1$, Sr214$-\delta\#1$, and Sr214$-\delta\#2$ at $T$ = 5 K with applied field perpendicular to the $c$-axis ($H$ $\parallel$ $ab$). Inset: expanded plot of $M$($H$) for Sr214$-\delta\#2$.
}
\end{figure}

Given that samples that were exposed to high temperature for longer time have lower sample quality, we further investigated the effect of prolonged exposure at 1300$^\circ$ C (Sr214$-\delta\#1$, see table \ref{table2}).
This sample showed significantly reduced magnetic ordering temperature, and a down turn in the $M(T)$ curve below $\sim$50 K. Most of the earlier data on Sr$_2$IrO$_4$ show such down turn in the $M(T)$ curve \cite{PhysRevB.84.100402,PhysRevB.80.140407}.
Our EDX measurements indicate that Sr214$-\delta\#1$ is deficient in oxygen (see table \ref{table1}).
Although EDX is not very sensitive to oxygen, the difference in the oxygen content from Sr214$\#1$ is statistically significant.
A similar magnetic behavior is observed in a sample (Sr214$-\delta\#2$) grown using the same heating sequence as in Sr214$\#1$ but with 10\% less molar ratio of SrCO$_3$. This further indicates that the different magnetic behavior is due to a non-stoichiometry in the sample, also supported by the EDX measurement of Sr214$-\delta\#2$ showing a further deviation from the ideal stoichiometry (table \ref{table2}).
In Fig.~\ref{FIGURE_AC} (a), we plot the $M$($T$) curves measured on zero field cooling (ZFC) and on field cooled warming (FCW) modes at $H$ = 5 kOe. Both Sr214$-\delta\#1$ and Sr214$-\delta\#2$ show substantial decrease in the magnetic moment and the transition temperatures ($T_N$).
Further, Sr214$-\delta\#2$ shows glassy behavior at low temperature.
In Fig.~\ref{FIGURE_AC} (b), we plot the inverse of the ZFC data to highlight the change in transition temperature from $T$ $\approx$ 240 K to $T$ $\approx$ 200 K.
For Sr214$-\delta\#1$ and Sr214$-\delta\#2$, $M$($H$) curves show hysteresis, which is absent in Sr214$\#1$ (Fig.~\ref{FIGURE_AC} (c)).

\begin{figure*}
\includegraphics[width=1\textwidth]{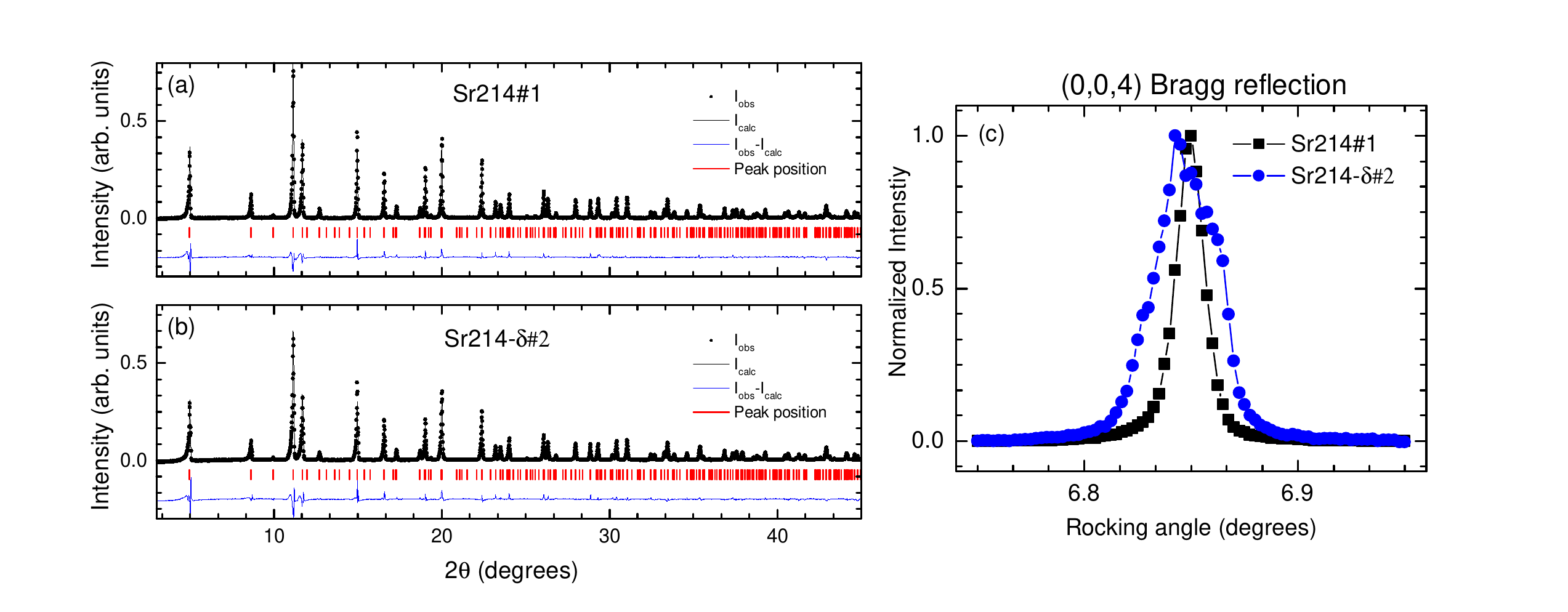}
\caption{\label{XRPD}
(Color online) X-ray powder diffraction pattern for (a) Sr214$\#1$ and (b) Sr214$-\delta\#2$ taken in Debye-Scherrer geometry using a 0.3mm diameter capillary. (c) Rocking curve through (0,0,4) reflection for single crystals of Sr214$\#1$ and Sr214$-\delta\#2$.
}
\end{figure*}

We now look for structural distortions associated with the oxygen vacancies.
Single crystal XRD experiments on Sr214$\#1$ and Sr214$-\delta\#2$ reveal an increase in the mosaicity of the non-stoichiometric sample as seen in the width of the rocking curve through the (0,0,4) Bragg reflection (Fig.~\ref{XRPD}(c)), pointing to decreased structural quality. Nevertheless, neither powder nor single crystal XRD resolve a systematic change in the lattice parameters or atomic positions, as seen in the nearly identical powder patterns in Fig.~\ref{XRPD}(a-b). Since x-rays are rather insensitive to oxygen atoms, we investigated the nature of this distortion by looking at the lattice dynamics using low temperature Raman spectroscopy.

Previous Raman results on Sr$_2$IrO$_4$ and Sr$_2$Ir$_{1-x}$Ru$_x$O$_4$ have observed that the 34 meV $A_{1g}$ phonon mode, associated with in-plane bending of the Ir-O-Ir bonds, is extremely sensitive to structural distortions \cite{PhysRevB.85.195148,PhysRevB.89.104406}.
An additional phonon mode was for instance found in the vicinity of this mode in Sr$_2$Ir$_{1-x}$Ru$_x$O$_4$ due to the formation of two Ir(Ru)-O-Ir(Ru) bond angles, indicating a phase separation \cite{PhysRevB.89.104406}.
We therefore look for signature of structural distortions in Sr214$\#1$, Sr214$-\delta\#1$, and Sr214$-\delta\#2$ by measuring their 34 meV in-plane bending mode with Raman scattering.

\begin{figure*}
\includegraphics[trim = 0mm 0mm 0mm 0mm, clip,width=1\textwidth]{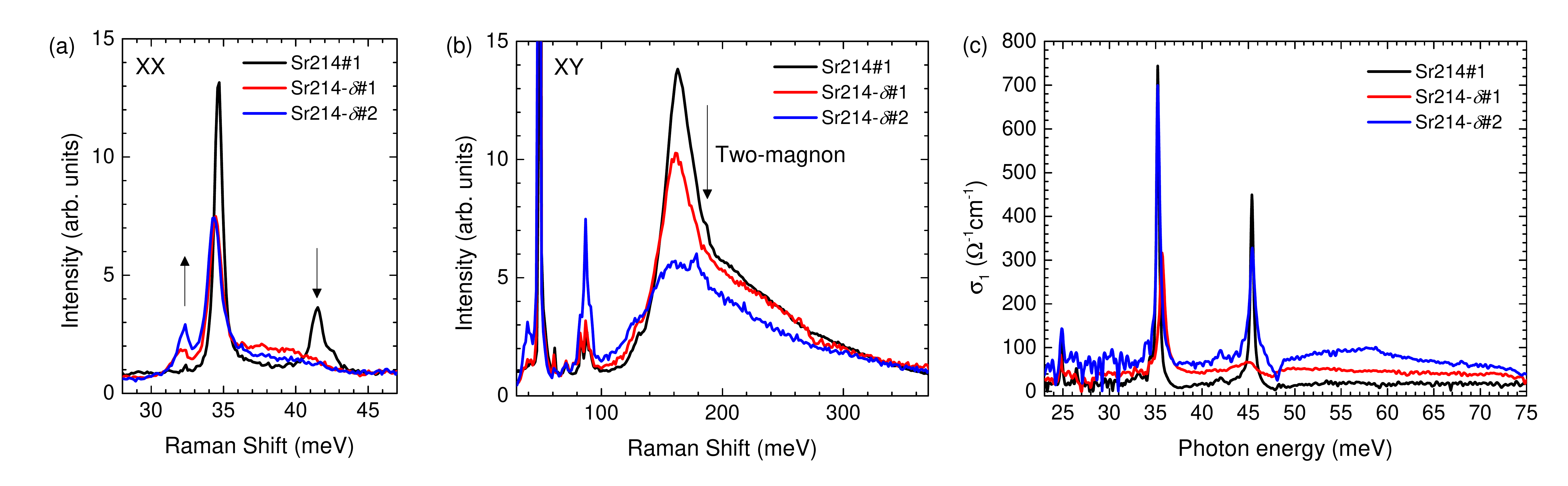}
\caption{
\label{Raman}(Color online) (a) Low energy Raman spectra of Sr214$\#1$, Sr214$-\delta$$\#1$, and Sr214$-\delta$$\#2$ taken at $T$ = 20 K
using the $XX$-channel ($A_{1g}$  and $B_{1g}$).
(b) High energy Raman spectra taken $T$ = 20 K using the $XY$-channel ($B_{2g}$) \cite{Gretarsson:Raman} and (c) real part of the FIR optical conductivity $\sigma_1(\omega)$  at $T=7\,\text{K}$ of the same samples.}
\end{figure*}

In Fig.~\ref{Raman} (a),  we show the low energy Raman spectra of Sr214$\#1$, Sr214$-\delta\#1$, and Sr214$-\delta\#2$ taken at $T$ = 20 K using the $XX$-channel probing modes with $A_{1g}$ and $B_{1g}$ symmetry.
Three $A_{1g}$ phonon modes are observed in the presented energy range for Sr214$\#1$, the aforementioned 34 meV mode along with two smaller modes around 42 meV. This is in good agreement with previous findings \cite{PhysRevB.85.195148}.
For Sr214$-\delta\#1$ and Sr214$-\delta\#2$, an additional Raman-active mode becomes visible around 32 meV, while the two modes around 42 meV vanish and the 34 meV mode broadens and shifts to lower energy.
This result is indicative of distorted in-plane Ir-O-Ir angle ($A_{1g}$) \cite{PhysRevB.89.104406}, most likely originating from in-plane oxygen vacancies.
In Fig.~\ref{Raman} (b), we show the high energy Raman spectra of the same samples taken $T$ = 20 K using the $XY$-channel ($B_{2g}$).
In this geometry the two-magnon signal can be observed \cite{Gretarsson:Raman}.
The two-magnon signal, found around 160 meV, becomes weaker and broader in Sr214$-\delta\#1$ and Sr214$-\delta\#2$.
These Raman results consistently indicate that oxygen vacancies lead to the observable difference in the magnetization data between Sr214$\#1$ and other samples.

We now investigate the effect of oxygen vacancies on the infrared (IR) active phonons and the in-gap spectral weight. Fig.~\ref{Raman} (c) shows the real part of the optical conductivity $\sigma_1(\omega)$ in the far-IR spectral range. Compared to the clean sample Sr214\#1 both oxygen-deficient samples show a significant increase in the background level of the optical conductivity $\sigma_1(\omega)$. We attribute this rise in absorption to an increased number of defect states due to oxygen vacancies.
While the phonon line shapes of sample {Sr214-$\delta$\#2} are rather similar to the clean case with just a small increase in the peak width the sample {Sr214-$\delta$\#1} shows a remarkable difference.
The phonon at $35.2\,\text{meV}$ experiences a significant energy shift to $35.7\,\text{meV}$ whereas the one centered at $45.4\,\text{meV}$ softens to $45.1\,\text{meV}$ and its intensity strongly decreases.
These two phonons have been assigned to in-plane bending modes of the $\text{IrO}_6$ oxygen octahedra \cite{PhysRevB.80.195110}. This suggests even stronger in-plane distortions of the oxygen octahedra in the {Sr214-$\delta$\#1} sample although the overall background level of absorption appears to be smaller compared to {Sr214-$\delta$\#2}.

\begin{table*}
  \centering
  \begin{tabular}{ | p{1.5cm} | p{2cm} | p{10cm} |}
    \hline
    Sr$_2$IrO$_4$ samples & IrO$_{2}$:SrCO$_{3}$ :SrCl$_{2}$ & Furnace sequence \\ \hline
    Sr214\#1  & 1 : 2 : 7 & 1300 $^{\circ}$C $\rightarrow$ (8 $^{\circ}$C/h) 900 $^{\circ}$C $\rightarrow$ RT \\ \hline
    Sr214\#2  & 1 : 2 : 7 &  1100 $^{\circ}$C $\rightarrow$ (45 $^{\circ}$C/h) 1300 $^{\circ}$C $\rightarrow$ (8 $^{\circ}$C/h) 900 $^{\circ}$C $\rightarrow$ RT \\ \hline
    Sr214\#3  & 1 : 2 : 7 &  1300 $^{\circ}$C (24h dwell)  $\rightarrow$ (8 $^{\circ}$C/h) 1100 $^{\circ}$C $\rightarrow$ Quench \\ \hline
    Sr214\#4  & 1 : 2 : 7 & 1300 $^{\circ}$C (24h dwell)  $\rightarrow$ (8 $^{\circ}$C/h) 1100 $^{\circ}$C (100h dwell) $\rightarrow$ RT\\
    \hline
    \end{tabular}

  \caption{Synthesis conditions for Sr$_2$IrO$_4$ crystals with fixed molar ratio of starting materials and various synthesis sequences.}\label{table1}
\end{table*}

\begin{table*}
  \centering
  \begin{tabular}{ | p{2cm} | p{2.5cm} | p{8cm} | p{3cm} |}
    \hline
    Sr$_2$IrO$_{4-d}$ samples & IrO$_{2}$:SrCO$_{3}$ :SrCl$_{2}$ & Furnace sequence & Chemical composition (EDX) \\ \hline
    Sr214\#1 & 1 : 2 : 7 & 1300 $^{\circ}$C $\rightarrow$ (8 $^{\circ}$C/h) 900 $^{\circ}$C $\rightarrow$ RT & Sr$_{2.08}$IrO$_{3.96}$\\ \hline
    Sr214$-\delta\#1$  & 1 : 2 : 7 & 1300 $^{\circ}$C (100h dwell) $\rightarrow$ RT & Sr$_{2.08}$IrO$_{3.86}$\\ \hline
    Sr214$-\delta\#2$  & 1 : 1.8 : 7 & 1300 $^{\circ}$C $\rightarrow$ (8 $^{\circ}$C/h) 900 $^{\circ}$C $\rightarrow$ RT & Sr$_{2.00}$IrO$_{3.68}$\\
    \hline
    \end{tabular}

  \caption{Synthesis conditions for stoichiometric Sr$_2$IrO$_4$ (Sr214$\#1$), non-stoichiometric Sr$_2$IrO$_{4-\delta}$ with oxygen vacancy (Sr214$-\delta\#1$ and $\#2$)}\label{table2}
\end{table*}

\begin{figure}
\includegraphics[width=0.5\textwidth]{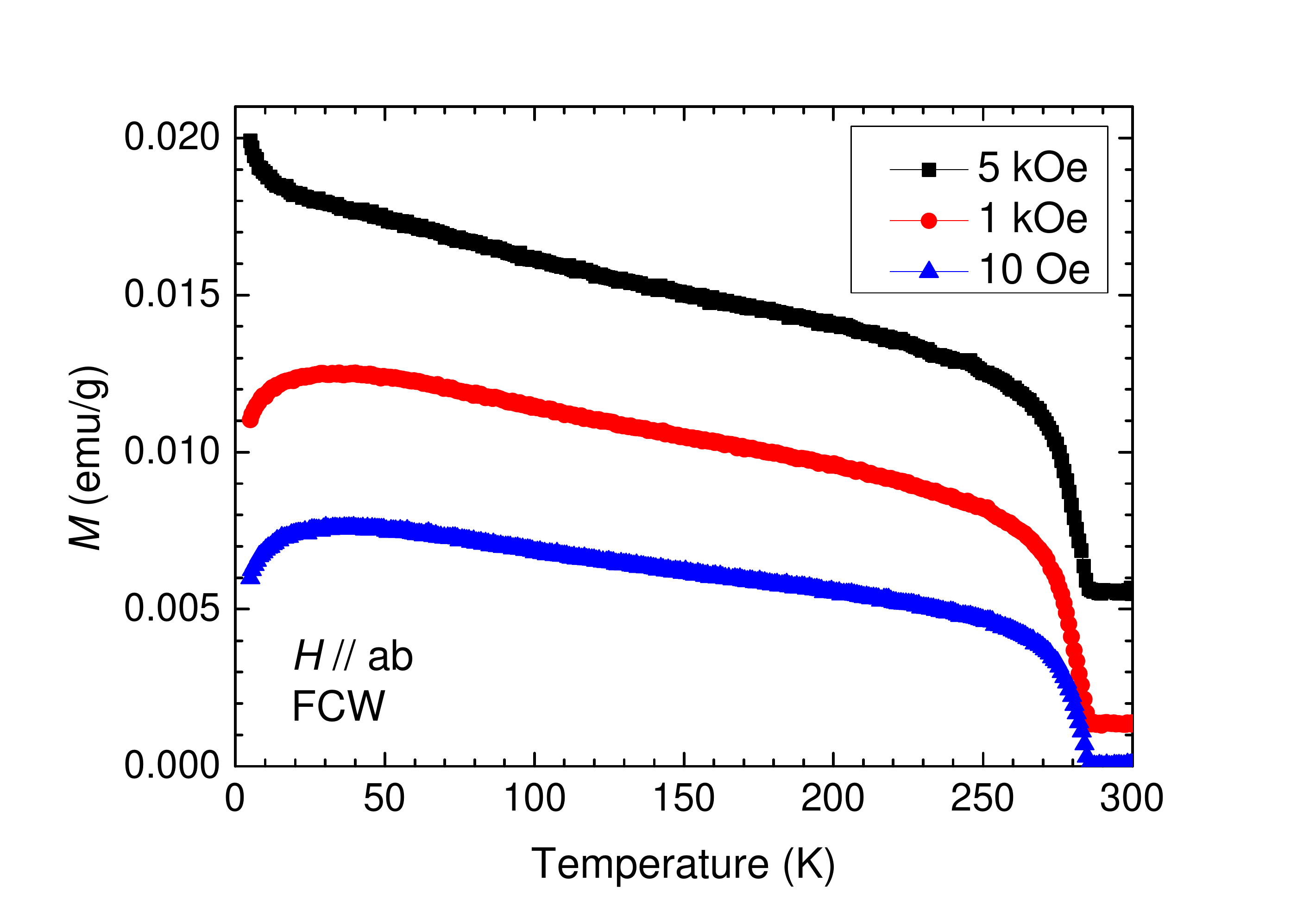}
\caption{
\label{FIGURE_H}(Color online) Temperature dependence of magnetization for a Sr$_3$Ir$_2$O$_7$ single crystal in field cooled warming (FCW) mode with applied field perpendicular to the $c$-axis ($H$ $\parallel$ $ab$) at $H$ = 10 Oe, 1 kOe, and 5 kOe.}
\end{figure}

\begin{table*}
  \centering
  \begin{tabular}{ | p{3cm} | p{2.5cm} | p{8cm} |}
    \hline
    Samples & IrO$_{2}$:SrCO$_{3}$ :SrCl$_{2}$ & Furnace sequence \\ \hline
    Sr214\#1     & 1 : 2 : 7 & 1300 $^{\circ}$C $\rightarrow$ (8 $^{\circ}$C/h) 900 $^{\circ}$C $\rightarrow$ RT \\ \hline
    Sr214/Sr327\#1   & 1 : 2 : 7 & 1150 $^{\circ}$C (12h dwell) $\rightarrow$ (5 $^{\circ}$C/h) 880 $^{\circ}$C $\rightarrow$ RT \\ \hline
    Sr214/Sr327\#2   & 1 : 2 : 7 & 1125 $^{\circ}$C (12h dwell) $\rightarrow$ (5 $^{\circ}$C/h) 880 $^{\circ}$C $\rightarrow$ RT\\ \hline
    Sr214/Sr327\#3   & 1 : 2 : 7 & 1050 $^{\circ}$C (12h dwell) $\rightarrow$ (5 $^{\circ}$C/h) 880 $^{\circ}$C $\rightarrow$ RT \\ \hline
    Sr327   & 2 : 3 : 7 & 1050 $^{\circ}$C (36h dwell) $\rightarrow$ (5 $^{\circ}$C/h) 750 $^{\circ}$C $\rightarrow$ RT \\
    \hline
    \end{tabular}

  \caption{Synthesis conditions for Sr$_2$IrO$_4$ (Sr214$\#1$), Sr$_3$Ir$_2$O$_7$ (Sr327) and intergrowth of Sr$_2$IrO$_4$ and Sr$_3$Ir$_2$O$_7$ (Sr214/Sr327\#1, \#2, and \#3).}\label{table3}
\end{table*}

\begin{figure}
\includegraphics[width=0.5\textwidth]{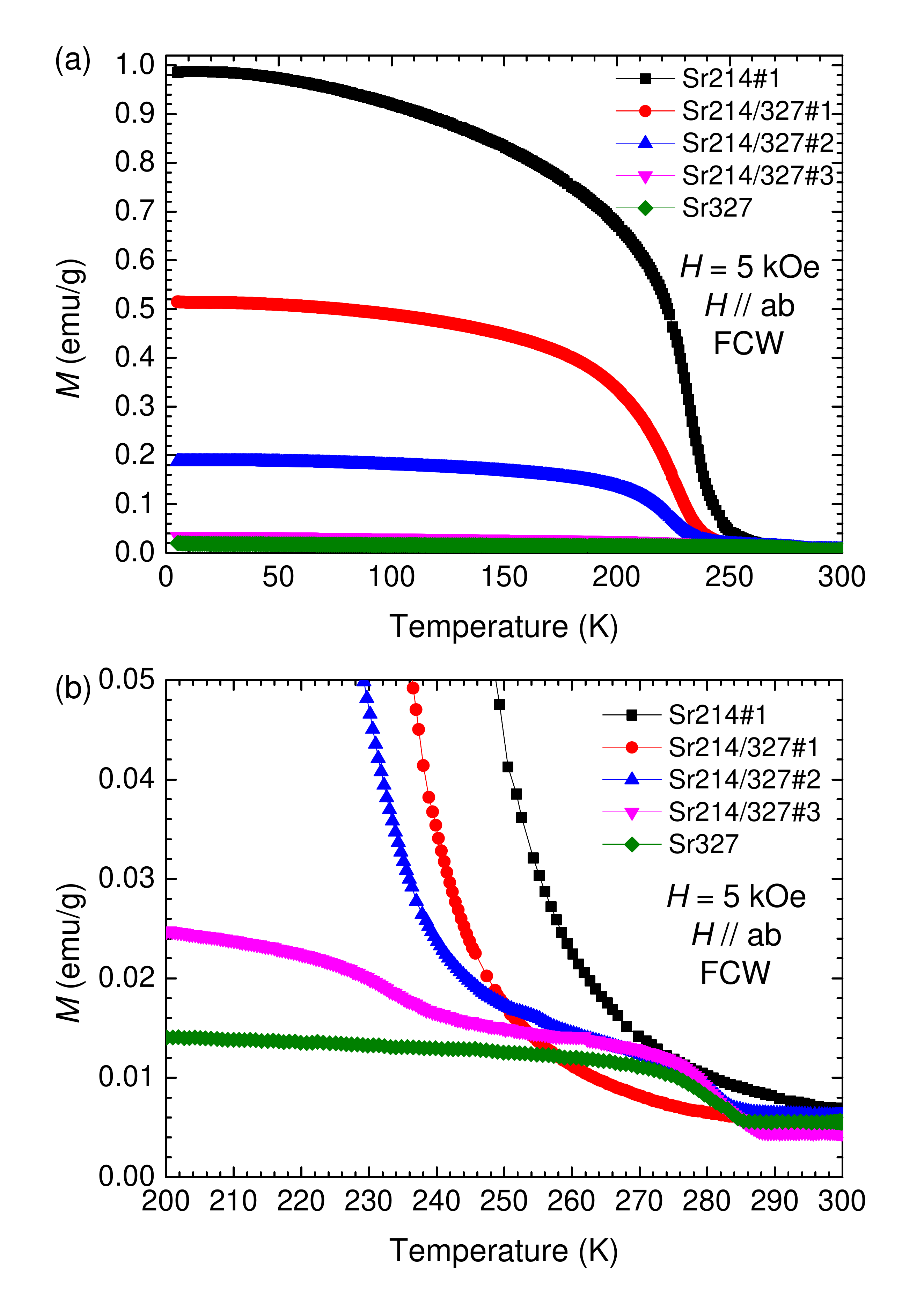}
\caption{
\label{FIGURE_D}(Color online) Temperature dependence of magnetization $M$($T$) at $H$ = 5 kOe ($H$ $\parallel$ $ab$) for Sr$_2$IrO$_4$ (Sr214$\#1$), Sr$_3$Ir$_2$O$_7$ (Sr327) and intergrowth samples (Sr214/327$\#1$, $\#2,$ and $\#3$).}
\end{figure}

We now move on to demonstrate that layers, or slabs, of Sr$_3$Ir$_2$O$_7$ can be introduced into a single crystal of Sr$_2$IrO$_4$ by further altering the growth sequence.
In table \ref{table3} the growth sequence and molar starting ratio of parent Sr$_2$IrO$_4$ and parent Sr$_3$Ir$_2$O$_7$ single crystals can be seen. We have confirmed the quality of our parent Sr$_3$Ir$_2$O$_7$ single crystals by measuring the magnetic susceptibility.
In Fig.~\ref{FIGURE_H}, we plot $M$($T$) for Sr$_3$Ir$_2$O$_7$ in FCW mode with applied field being perpendicular to the $c$-axis ($H$ $\parallel$ $ab$). Three different fields were used, $H$ = 10 Oe, 1 kOe, and 5 kOe. In all measurements a magnetic phase transition is observed at $T$$_N$ $\approx$ 285 K. At lower temperatures, no other transition is seen in contrast to previous results \cite{PhysRevB.66.214412, PhysRevB.85.184432}.

One can notice that in order to produce pure Sr$_3$Ir$_2$O$_7$ single crystals the starting temperature of the growth sequence for Sr214$\#1$ has to be substantially reduced, along with a change in the molar ratio.
Therefore one way to induce intergrowth of Sr$_2$IrO$_4$ and Sr$_3$Ir$_2$O$_7$ would be to systematically reduce the starting temperature of the growth sequence for Sr214$\#1$, from 1300 $^{\circ}$C to 1050 $^{\circ}$C, while at the same time keep the molar starting ratio unchanged. This approach is displayed in table \ref{table3} where we present in details the growth sequence for intergrowth of Sr$_2$IrO$_4$ and Sr$_3$Ir$_2$O$_7$ samples (Sr214/Sr327$\#1$, $\#2$, and $\#3$).

\begin{figure}
\includegraphics[width=0.5\textwidth]{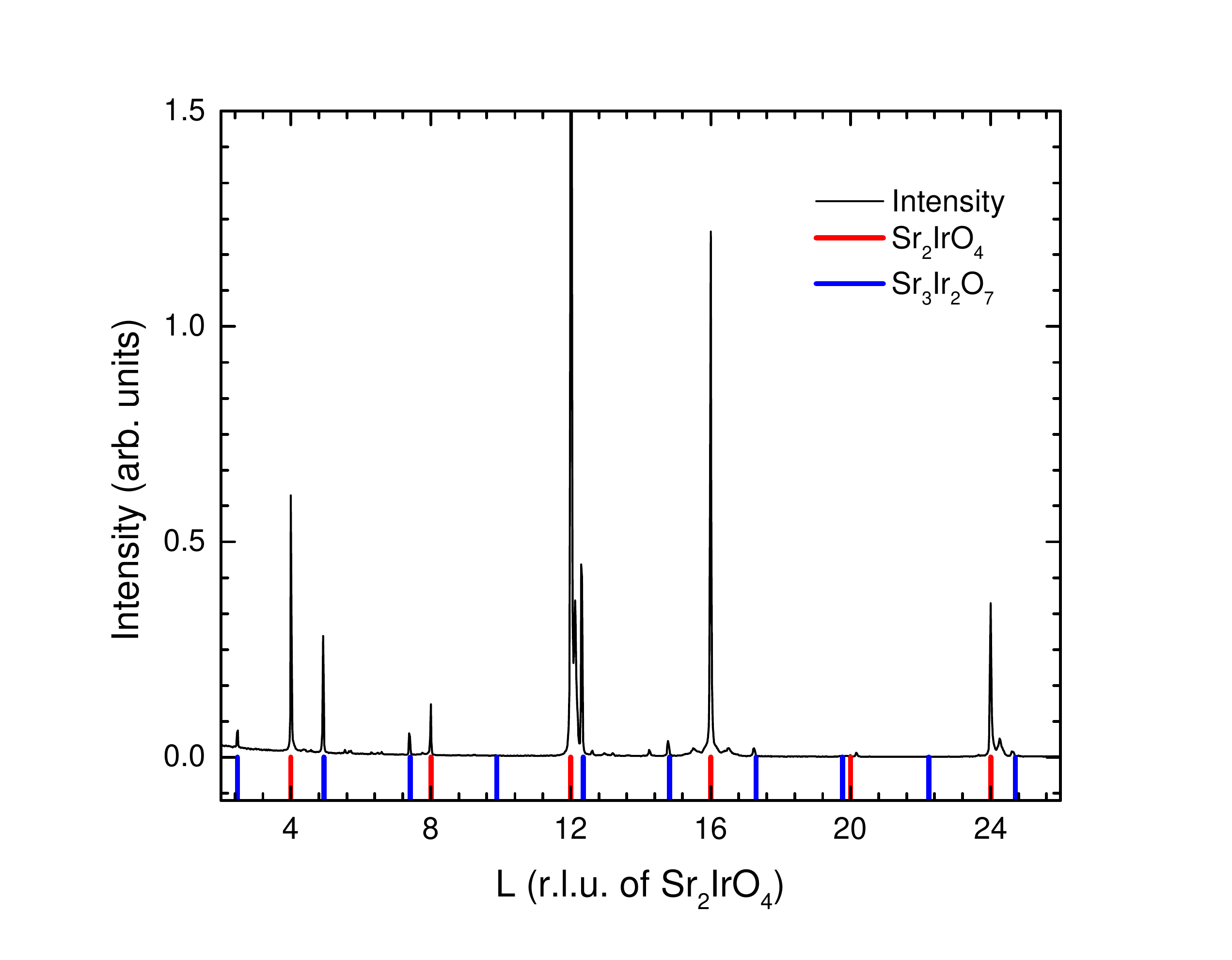}
\caption{\label{XRD214_327}
(Color online) Single crystal x-ray diffraction pattern of the (0,0,L) reflections for a crystal grown using similar conditions to Sr214/Sr327$\#1$. Peaks identified with the periodicities of Sr$_2$IrO$_4$ and Sr$_3$Ir$_2$O$_7$ are denoted by vertical lines. Small peaks in the background are associated with stacking faults.
}
\end{figure}

In Fig.~\ref{FIGURE_D} (a) we plot the temperature dependence of the magnetization $M$($T$) of each crystal in table \ref{table3} with $H$ $\parallel$ $ab$. Comparison of different curves reveals that by lowering the starting temperature of the growth sequence the saturated magnetic moment gradually approaches the value measured for pure Sr$_3$Ir$_2$O$_7$. Interestingly, in Fig.~\ref{FIGURE_D} (b), where we show a closer look at the onset of magnetic order,  one can notice a two-step structure in the $M$($T$) curve of Sr214/Sr327$\#2$ and $\#3$. Comparison to pure Sr$_3$Ir$_2$O$_7$ reveals that the transtion at $T$$_N$ $\approx$ 285 K is associated with the Sr$_3$Ir$_2$O$_7$ portion of the intergrowth samples while the stronger transition $T$$_N$ $\approx$ 240 K comes from the Sr$_2$IrO$_4$ part. Moreover, by looking at the saturated magnetic moments in $M$($T$) (see Fig.~\ref{FIGURE_D} (a)), the ratio of the two phases can be estimated; Sr$_2$IrO$_4$ has a saturation magnetic moment of $\sim$1 emu/g while in comparison, in Sr$_3$Ir$_2$O$_7$ the magnetic moment at $H$ = 5 kOe is negligible ($H$ $\parallel$ $ab$). Using this method we can control, by varying the synthesis conditions (table \ref{table3}), the proportion of the two phases. The Sr$_2$IrO$_4$ contribution to Sr214/Sr327$\#1$ is about 0.5 while this ratio drops to about 0.2 in Sr214/Sr327$\#2$ and around 0.03 in Sr214/Sr327$\#3$. Given the difference in the $c$-axis lattice parameter, the intergrowth nature of these samples is confirmed by using single crystal XRD to study the Bragg reflections along L. As seen in Fig.~\ref{XRD214_327}, reflections corresponding to Sr$_2$IrO$_4$ or Sr$_3$Ir$_2$O$_7$ can be observed in the same single crystal.

\section{Conclusion}

We have investigated the relationship between different synthesis conditions and magnetic properties in Sr$_2$IrO$_4$ single crystals.
Depending on the synthesis condition we observe varying magnetic properties most likely due to oxygen vacancies.
For off-stoichiometric samples, additional phonon-modes appear in Raman scattering spectra due to structural distortions and in-gap states appear in optical conductivity measurements. Consistent with the magnetization measurements, we find that the two-magnon Raman peak broadens, signaling weaker magnetic correlations in these distorted samples.
In addition, we have shown that intergrowth of the Sr$_3$Ir$_2$O$_7$ phase into Sr$_2$IrO$_4$ single crystals can be be controlled by varying the synthesis conditions. Our results therefore emphasizes that revealing the intrinsic physical properties of Sr$_2$IrO$_4$ and related materials require scrupulous control of the synthesis conditions.

\acknowledgments
N. H. Sung was supported by the Alexander von Humboldt Foundation. We gratefully acknowledge Y.-L.~Mathis for support at the IR1 beam line of the synchrotron facility ANKA at the Karlsruhe Institute of Technology. We acknowledge financial support of the DFG under SFB/TRR80.

{}

\end{document}